\documentstyle[aps,twocolumn,prl,epsf]{revtex}

\renewcommand{\ref}[1]{\raisebox{.6ex}{[#1]}}

\newcommand{\be}{\begin{equation}}
\newcommand{\ee}{\end{equation}}

\newcommand{\ba}{\begin{array}}
\newcommand{\ea}{\end{array}}

\newcommand{\bea}{\begin{eqnarray}}
\newcommand{\eea}{\end{eqnarray}}

\begin{document}


\twocolumn[
\hsize\textwidth\columnwidth\hsize\csname @twocolumnfalse\endcsname

\title{ Role of Impurities in Core Contribution to Vortex Friction }

\author{ P. Ao$^a$ and X.-M. Zhu$^b$ \\ 
      Department of Theoretical$^a$/Experimental$^b$ Physics  \\
     Ume\aa{\ }University, S-901 87, Ume\aa, SWEDEN  }

\maketitle

\widetext

\begin{abstract} 
We present a microscopic calculation 
of  core states contribution to vortex friction
in a type II superconductor at zero temperature. 
In weak impurity limit, a perturbative calculation
leads to a  vortex friction proportional to the normal state resistivity. 
In dirty limit,  vortex friction due to
 core states is determined only by 
the Fermi energy and the energy gap, 
similar to the  phenomenological 
results of Bardeen and Stephen,
but disagrees with the results obtained under the relaxation time  
approximation.  In addition, 
an explicit demonstration is given on the
the insensitivity of  transverse force to impurities.

\end{abstract}

\noindent
PACS \#'s: 74.60.Ge; 47.37.+8; 67.40.Vs

] 

 
\narrowtext


The importance of vortex dynamics in a superconductor
has long been realized\cite{bs,nv} and been
under continuous and intensive studies\cite{kim}. 
In the early work by Bardeen and Stephen, it was found
that in  dirty limit and at
 zero temperature, vortex friction is 
determined by  the Fermi energy and  
the energy gap of the superconductor.\cite{bs}
This phenomenological result has ample experimental 
support\cite{kim}.
Subsequent microscopic derivations,
however, have reached a conclusion that
 the vortex friction should approach  zero in dirty limit.
Meanwhile, these calculations also give vanishing
a transverse force on a moving vortex in the same limit.\cite{russian}
 
The magnitude of  transverse 
force on a moving vortex has attracted 
much attention in recent years.
The Berry phase acquired by a moving vortex
shows that the Mangus force is the  
only transverse on the vortex at zero
temperature irrespective  whether or not impurities are presented.\cite{at}   
This observation has been confirmed by a 
direct measurement of the transverse
force in dirty superconductors\cite{zhu}.
It has led to a re-examination of the
approximations used in the earlier
microscopic derivations of vortex dynamics.
It has been found that the relaxation time approximation
always leads to incorrect results
when used in force calculations by either
force correlation or force balance conditions.\cite{ao}
The calculations which give  a 
vanishing vortex friction in dirty limit, in
disagreement with the Bardeen-Stephen result,
are exactly those using relaxation time approximations.

The problem now becomes how to calculate
vortex friction 
from a microscopic theory in presence of impurities without
the improper use of the relaxation time approximation.
Solving this problem is the main purpose of our paper.
In the present  paper we will
report a set of results based on the BCS theory
which shows that  core states   
contribution to vortex friction arises and saturates
as the superconductor evolves from weak impurity to  dirty limit,
determined by the energy spectral property of the BCS Hamiltonian.
An explicit demonstration of the insensitivity of the transverse force
to impurities will also be given here, providing another way to 
concretely realize the results from the Berry phase calculation.

We consider an isolated rectilinear vortex at $x_v$ of length $L$ 
in a BCS superconductor. For our purpose of calculating
core states contribution to vortex dynamics, we will not
include the vector potential of a magnetic field explicitly
because its effect on the core state structure is negligible.
The vortex velocity is given by  ${\bf v}_v =  \dot{x}_v$. 
The distribution of impurities is assumed to 
be homogeneous and the variation of impurity potential only
appreciable on a scale much smaller than size of the vortex core.
If the vortex is allowed to move slowly against the ionic lattice 
background, 
to the leading order of the vortex velocity, 
according to the Ehrenfest theorem, 
the corresponding force acting on  the vortex has the form\cite{tan}
$
   {\bf F} = B {\bf v}_v \times \hat{\bf z} -  
            \eta {\bf v}_v  \; .
$
Within the BCS theory, 
the transverse coefficient $B$ of the transverse force is
explicitly expressed 
in terms of the transverse force-force correlation function\cite{az,za} 
\bea
    B & = &  \frac{i \hbar }{2} 
    \sum_{k,k'} \frac{ f_k - f_{k'} }{ ( E_k - E_{k'} )^2 } \int d^3x d^3x' 
          \times \nonumber \\
    & & \left( \Psi_k^{\dag}(x')\nabla_v {\cal H} \Psi_{k'}(x')\times
      \Psi_{k'}^{\dag}(x)\nabla_v {\cal H} \Psi_{k}(x) \right) 
        \cdot \hat{\bf z}   \; ,
\eea
and the vortex friction coefficient $\eta$ in terms of the longitudinal 
force-force correlation function
\bea
    \eta & = & \frac{\pi }{4}
     \sum_{ k'\ne k }   \hbar  
     \frac{f_k - f_{k'} }{ E_k- E_{k'} }
  \delta( 0^{+} - |E_{k'}- E_k|  ) \times  \nonumber \\
  & &   |\langle\Psi_k|\nabla_v {\cal H} |\Psi_{k' } \rangle |^2  \, . 
\eea 
Here  $f_k = 1/(1 + e^{\beta E_k} ) $ is the Fermi distribution function.
The wavefunctions $\{ \Psi_k(x) \}$ and the corresponding
eigenvalues $ \{ E_k \} $  are determined by 
the Bogoliubov-de Gennes equation,
$
   \ba{l}
   {\cal H} \; \Psi_k(x) = E_k \; \Psi_k(x) \; ,
   \ea
$
with 
$
   \Psi_k(x) = \left( \begin{array}{c} u_k(x) \\ v_k(x) \end{array} 
       \right) \; ,
$   
and the system Hamiltonian given by 
$
  {\cal H} = \left(\begin{array}{cc} 
                     H  & \Delta  \\
                    \Delta^{\ast} & - H^{\ast} 
                   \end{array} \right) \; .
$
Here $H =  - (\hbar^{2}/2m ) \nabla^{2} -  \mu_F + V(x) $, and
$V(x)$ the impurity potential.
The order parameter is determined self-consistently
\[
  \Delta(x) = g < \sum_k (1-2f_k) u_k(x) v_k^{\ast}(x) > \; , 
\]
where the impurity average $< ... > $ is implied.
The amplitude of the order parameter $|\Delta|$ is the energy gap,
and  $g$ the strength of attractive interaction between electrons.
Since the impurity distribution is homogeneous, its average leaves the
vortex position the only reference point in the system. 
We expect $\Delta(x,x_v) = \Delta ( x - x_v) 
= |\Delta(x-x_v)| e^{i\theta(x-x_v) } $ with $\theta$ the singular phase. 
This gives a convenient specification of the 
vortex position in the calculation, which we will adopt in this paper.
 

We will  first evaluate the transverse force to show its insensitivity 
 to impurities. 
We   demonstrate that  
$B$ can be expressed in a manner without explicit dependence on
 the system Hamiltonian. 
Because $ {\cal H} $ is Hermitian, all its eigenstates form a  
complete and orthonormal set, that is, 
$
      \int d^3 x \Psi^{\dag}_k(x)\Psi_{k'}(x)  
    = \delta_{k,k'} \; ,
$
and 
$
  \sum_k \Psi_k(x)\Psi_k^{\dag}(x')           
     =  {\bf 1}  \; , 
$
with $\Psi^{\dag}_k(x) = ( u^{\ast}_k(x), v^{\ast}_k(x) ) $ and the 
wavefunction
is normalized to 1 over a cylinder of radium $R$ and length $L$.
The thermodynamic limit requires that $R \rightarrow \infty$.
For states $k\neq k'$ we have 
\bea 
    & & \int d^3x  \; \Psi_k^{\dag}(x)  (\nabla_v {\cal H} )\Psi_{k'}(x) 
     / (E_{k'} - E_{k} )   \nonumber   \\
    & = &  \int d^3x \; \Psi_k^{\dag}(x)  
      \nabla_v \Psi_{k'}(x) 
     =  - \int d^3x \; \nabla_v \Psi_k^{\dag}(x)  
     \Psi_{k'}(x)  \; ,
\eea
where a detailed solution of these states  from 
the  Bogoliubov-de Gennes equation with a vortex 
can be found in Ref.\cite{bardeen}.
Eq.(3) is obtained by taking gradient $ \nabla_v$ with respect to
$ {\cal H} |\Psi_{k'} \rangle =  E_{k'} |\Psi_{k'} \rangle$
and $   \langle \Psi_k|{\cal H} =  E_k \langle \Psi_k|$,
then multiplying  from left or right by 
$ \langle \Psi_k|$ or $ |\Psi_{k'} \rangle $ respectively.
With the aid of Eq.(3) and the completeness
of the eigenfunctions, the transverse 
coefficient can be expressed as a summation over  individual state 
contributions,
\bea
   B &=& i\hbar \sum_{k} \int d^3 x 
     \left\{ f_k \nabla_v u_k(x)\times \nabla_v u^{\ast}_k(x)  \right.
           \nonumber \\
    &&   \left.  -  (1-f_k) \nabla_v v_k(x) \times \nabla_v v_k^{\ast}(x)  
        \right\}\cdot \hat{\bf z} \; . 
\eea
Now we are ready to explicitly demonstrate that 
the coefficient $B$ for the transverse force 
is insensitive to impurities as
first observed through the Berry phase calculation.

In a clean superconductor we can use the replacement of 
$\nabla_v \rightarrow -\nabla$ in Eq.(4). 
Using this replacement and the definition of number current,
\[ 
  {\bf j} = - \frac{i}{2} \sum_k \left\{ f_k u_k^{\ast} \nabla u_k +
  (1-f_k) v_k \nabla v_k^{\ast} \right\} + c.c.  \; ,
\] 
Eq.(4) becomes
\bea
  B & = & \hbar \int d^3 x \; \hat{z}\cdot(\nabla\times{\bf j}) \nonumber \\
    & = & L \hbar \oint_{|x-x_v| \rightarrow \infty}  
                  d{\bf l} \cdot{\bf j} \nonumber  \\
    & = & L\; 2\pi\hbar\rho_s(T)    \; . 
\eea
At zero temperature, $f_k = 1 (0)$ for $E_k < 0 (> 0)$, 
 the calculation of $\rho_s$ is 
straightforward. It is equal to the total superfluid number density 
$\rho_0 =  \sum_{k, E_k > 0 } |v_k(|x - x_v| \rightarrow 
    \infty)|^2 $, number of Cooper pairs per unit area.
Here we have used the relations between the positive and negative
energy states of the Bogoliubov-de Gennes equation:
If 
$
    {\cal H} \;  \Psi(x) = E \; \Psi(x)    \; ,  \;    
 \overline{\Psi}(x) = \left( \begin{array}{c} 
                                v^{\ast}(x) \\ 
                      - u^{\ast}(x)\end{array} \right) \; ,
$
then
$
     {\cal H} \; \overline{\Psi}(x) = - E \; \overline{\Psi}(x) \; .
$ 

In the presence of impurities, the replacement
$\nabla_v \rightarrow -\nabla$ cannot be directly used in Eq.(4)
because of the implicit impurity dependence.
Instead, we expand $\Psi_k$ in terms of eigenfunctions
 $\{ {\Phi}_l \} $  of $\bar{\cal H}$, 
a corresponding  Hamiltonian
to ${\cal H}$  without the impurity potential $V(x) $ and with impurity 
averaged $\Delta$, 
\be
    \Psi_{k} = \sum_{ l } a_{k l}e^{i\delta_{k l}} \; \Phi_{l} \, .
\ee
Here
$ \{ a_{k\mu} \} $ and   $ \{ \delta_{k\mu} \} $ 
are the modulus and phases of the expansion coefficients.
They are functions of the vortex position 
as well as the positions of impurities.
We assume these coefficients to be 
described separately by two independent random matrices,
making use of the randomness of impurities.\cite{random}
$\bar{\cal H}$ still has a dependence on impurities through  the 
impurity averaged order parameter $\Delta$:  
\be
   \bar{{\cal H} } = \left(\begin{array}{cc} 
                     \bar{ H } & \Delta  \\
                    \Delta^{\ast} & - \bar{H} ^{\ast} 
                   \end{array} \right) \; ,
\ee
with $\bar{H} =  - (\hbar^{2}/2m ) \nabla^{2} -  \mu_F $,
$\bar{\cal H} \Phi_l = \bar{E}_l \Phi_l $
and 
$ {\Phi}_l = \left( \ba{l} \bar{u}_l \\ \bar{v}_l 
                      \ea \right) $ . 
Since $ \{ \Phi_\mu \} $ form a complete set, the expansion coefficients
$ \{a_{k l} e^{i\delta_{k l} }  \}$ 
form a unitary matrix, and $\sum_{l \, or \, k} a_{kl}^2 = 1 $.
For ${\Phi}_l $ we can use the replacement $\nabla_v \rightarrow - \nabla$.
We remind here that  
although  away from the vortex core the value of the 
order parameter is the same  as that in the clean case, guaranteed by
the Anderson theorem,  
in the core regime the profile of  $\Delta$ is different,
corresponding to a different coherence length.
The core size will be smaller with impurities presented.
Using Eq.(6), 
%
Eq.(4) becomes
\bea
   & & B  =  - i\hbar 
      < \sum_k \sum_{l,l'} \int d^3 x  
           \left\{ f_k 
       \nabla_v \left( a_{k l} e^{i \delta_{k l} } \bar{u}_{l}(x) 
         \right) 
        \times   \nonumber \right.   \\
   & & 
       \nabla_v \left( a_{k l'} e^{ - i \delta_{k l'} } 
         \bar{u}_{l'}^{\ast} (x)  \right)  
             - ( 1 - f_k )
       \nabla_v \left( a_{k l} e^{i \delta_{k l} } \bar{v}_{l}(x) 
         \right) \times      \nonumber \\
   & &  \left.  
       \nabla_v \left( a_{k l'} e^{ - i \delta_{k l'} } 
         \bar{v}_{l'}^{\ast} (x)  \right) \right\} \cdot \hat{\bf z} > \; .
\eea
Here $< ... > $ stands for the impurity average 
over the expansion coefficients. 
Then,
\bea
   B & = & - i\hbar  
       \sum_k \sum_{l}  <a_{k l}^2 > \int d^3 x  \left\{ f_k 
        \nabla_v\bar{u}_l(x) 
        \times  
       \nabla_v \bar{u}_{l}^{\ast}(x)  \right. \nonumber \\
   & &   \left.   -   (1-f_k) 
        \nabla_v \bar{v}_l(x) 
        \times  
       \nabla_v \bar{v}_{l}^{\ast}(x)  \right\} \cdot \hat{\bf z} \; .
\eea
Terms containing derivative to vortex position 
in the expansion coefficients have been averaged to zero. 
Now the replacement of $\nabla_v \rightarrow - \nabla $ can be used, and 
following the procedure leading to Eq.(5) we have, at zero temperature,
the desired result
$
 B  = 2 \pi \hbar \rho_0 \; ,
$
because 
\bea
  & & \sum_{k, E_k > 0 } \sum_{l} < a_{k l}^2 >
       | \bar{v}_l(|x-x_v| \rightarrow \infty )|^2 \nonumber \\  
  & = & \sum_{k, E_k > 0 } < |v_k(|x-x_v | \rightarrow \infty )|^2 >
         = \rho_0  \; .
\eea        
The above second equality is the Anderson theorem. 
Eq.(10) can also be reached from the envelop 
wavefunction argumentation\cite{degennes}.  

In the following we turn to the calculation of  
the core states contribution  to vortex friction at zero temperature.
We start with a clean superconductor  at zero temperature. 
In this limit core state energies are completely labeled 
monotonically 
by a half integer $\mu$ associated with the azimuthal angle $\theta$ in 
the eigenfunction $\Psi_l$\cite{bardeen}and the core 
energy spectrum is discrete.
In the transition matrix element shown in Eq.(2), 
only those between neighboring $\mu$
 are non-zero. 
In addition, because of the factor $ (f_k - f_{k'})$ in Eq.(2),
 only the two states closest to the Fermi surface, 
one above and one below, need to be considered. 
However due to  the finite energy level spacing, 
their contribution to the vortex friction is zero, because of the  
delta function in energy in Eq.(2).
Therefore, at zero temperature and 
in the clean limit there is no core states 
contribution to  vortex friction.

In the presence of a finite impurity potential $V(x)$, the energy spectrum 
of core states no longer has a monotonic dependence on $\mu$.
Transition elements in Eq.(2) are no longer confined to 
neighboring $\mu$ states.
The impurity potential also brings eigenenergies of some of the core states 
close to each other so that it generates
a quasi-continuous energy spectrum after impurity average, necessary for 
a finite vortex friction  according to Eq.(2).
In the weak impurity potential limit, their contribution to
the vortex friction can be calculated perturbatively. 
We will show that vortex friction due to core states is 
 proportional to the normal state resistivity.

To explicitly exploit the perturbation method, it is more convenient to
use $ \nabla \Psi_{k}(x)$ instead of $\nabla_v \Psi_{k}(x) $ in Eq.(2). 
It may be done in the following manner. 
The system Hamiltonian may be written as 
${\cal H} = \bar{{\cal H} } + {\cal  H}' $, with 
  $ \bar{{\cal H} } $ defined by  Eq.(7), 
and  
\[
   {\cal H}' = \left( \begin{array}{cc} 
                     V(x) & 0   \\
                    0  & - V(x) 
                   \end{array} \right) \; .
\]
Next we use
\[
  (\nabla {\cal H} ) 
   = \nabla \bar{\cal H}   + \nabla {\cal H}'  
   = - ( \nabla_v \bar{\cal H} )  + \nabla {\cal H}'  
   = - ( \nabla_v  {\cal H} )  + \nabla {\cal H}'  \; ,
\]
by noting that  $  \bar{\cal H} $ and  $ \nabla_v  {\cal H} $
have same dependence on  $x_{v} $, in addition to
\[ 
  \int d^3x \; \Psi_k^{\dag}(x)  \nabla ( {\cal H} \Psi_{k'}(x) ) 
\]
\[
  = \int d^3x \left[ 
    \Psi_k^{\dag}(x) ( \nabla {\cal H} ) \Psi_{k'}(x) 
    + \Psi_k^{\dag}(x)   {\cal H}  \nabla  \Psi_{k'}(x)\right] \, , 
\]
 and Eq.(3) to obtain
\bea
    & &  \int d^3x \; \Psi_k^{\dag}(x)  
      (\nabla_v {\cal H} ) \Psi_{k'}(x) = 
    - (E_{k'} - E_{k} ) \times \nonumber \\ 
   & &  \int d^3x \; \Psi_k^{\dag}(x)  
      \nabla \Psi_{k'}(x) + \int d^3x \; \Psi_k^{\dag}(x)  
      \nabla   {\cal H}'\Psi_{k'}(x) \, .
\eea
The effect of impurities is now described by $\nabla {\cal H}'$.
With Eq.(11), 
the transition element is given by
\bea
    & & < \left|\int  d^3x \; \Psi_k^{\dag}(x)
      \nabla_v {\cal H} \Psi_{k'}(x) \right|^2  >   = 
  <  \left| - (E_{k'}-E_k) \int d^3 x \times  \right.  \nonumber \\
 & & \left.  \Psi_k^{\dag}(x)\nabla \Psi_{k'}(x)
     + \int d^3x \; \Psi_k^{\dag}(x)\nabla{\cal H}' 
       \Psi_{k'}(x) \right|^2 > \, .
\eea
For the leading order of contribution, we 
use the unperturbed $\Psi_k$ and $\Delta$. 
The first term and the cross terms in Eq.(12) will not give any contribution 
to the vortex friction because of the discreteness of $(E_{k'}-E_k)$ and
the  delta function factor
$ \delta( E_k - E_{k'}  )$ in the vortex friction coefficient $\eta $. 
The contribution to $\eta$ comes from the last term.
Without losing  generality, we assume that the impurity 
potential has a length scale smaller than the coherence length
such that we may effectively describe it by a delta 
potential $V(x)=\sum_i V_0 \delta^3 (x-x_i)$, 
with $V_0$ carrying a random sign and $x_i$ randomly distributed.  
Using the wavefunction obtained in Ref.\cite{bardeen}, we obtain
the last  term of Eq.(12), 
\[
 < \left| \int d^3x \Psi_k^{\dag}(x)\nabla {\cal H}' \Psi_{k'}(x) 
   \right|^2 >
  \approx n_i (\pi\xi^2 L) V_0^2 \left(\frac{k_F}{\pi\xi^2 L}\right)^2  \; .
\]
Here $n_i$ is the impurity concentration.
Hence the vortex friction coefficient is 
\be
   \eta =  \frac{\pi}{2} {\hbar } 
    \left( \frac{ k_F L }{2\pi \epsilon_0} \right)^2 
    n_i (\pi\xi^2 L) V_0^2 \left(\frac{k_F}{\pi\xi^2 L}\right)^2  \; .
\ee
Here $\epsilon_0 = \Delta^2_\infty /E_F$ is the core level spacing,
$\xi k_F = E_F/\Delta_\infty$, 
$E_F = (\hbar k_F)^2/2m$, and $\Delta_\infty$ is the energy gap far away 
from the vortex core. 
In Eq.(13)
 $ k_F L /(2\pi{\epsilon_0} ) = n_c(E) $ is the approximated  number of core 
states per unit energy near the Fermi surface,  
arose from  the summation
\bea
    & & \sum_{k,k'} \delta( E_k - E_{k'} ) 
     \frac{ f_k - f_{k'} }{E_{k'} - E_k } \nonumber \\
    & = & \int d E_k d E_{k'}  \; \delta(E_k - E_{k'})
    \frac{ f_k - f_{k'} } {E_{k'} - E_k }   \,  n^2_c(E)
      =  n^2_c(E)  \nonumber \;  .
\eea
We now make connections to the normal states
transport coefficients.  Note that  for the normal state the electronic 
relaxation time and the electron scattering cross section have the
following relations, 
$
 \tau^{-1}_{tr} =  n_i  v_F \sigma_{tr} \; ,
$
with the transport cross section
$
 \sigma_{tr} = \int d\Omega \; ( 1 - \cos\theta)|V(\theta)|^2  \; .   
$
\cite{mahan}
Here
$
  V(\theta) = -{m}/{2\pi\hbar^2} \; \int d^3x V(x)
   e^{-i{\bf q}\cdot{\bf  r}} \, , 
$
with ${\bf q}= {\bf k} - {\bf k'}$, $\theta$ is the angle
between ${\bf k}$ and ${\bf k'}$, and $v_F = \hbar k_F/m$.
For our choice of potential, 
$
   \tau^{-1}_{tr} =  n_i v_F
   ({m}/{2\pi\hbar^2})^2  V_0^2 
   \; 4\pi  \, .
$
Expressed in $\tau_{tr}$ for the transition element, 
the vortex friction coefficient becomes 
\be
   \eta = \frac{3}{2} \frac{m n_e (\pi\xi^2 L)}{\tau_{tr}} \; ,
\ee
with the electron density $n_e = k_F^3 / (3\pi^2)$ . 
The scattering time $\tau_{tr}$ may be related to  the
normal state residual resistivity by 
$
  \rho = {m}/{ (n_e e^2 \tau_{tr}) } 
$,
and it can be measured independently.
Eq.(14)  shows that the vortex friction in the weak impurity limit
increases with the normal state resistivity.
We show below that this increase eventually saturates in the dirty limit.

Using  Eq.(6),
we expand localized states $  \Psi_k$ in terms of $\{ \Phi_l \}$, 
the eigenfunctions of $\bar{\cal H} $. Because 
$\nabla_v {\cal H} = \nabla_v \bar{\cal H }$,
\bea
   & & < \left |\int d^3x \; \Psi_k^{\dag}(x) \left(\nabla_v {\cal H}\right) 
            \Psi_{k'}(x) \right|^2  >
     \nonumber \\
   & & =  \sum_{l_1, l_1', l_2, l_2'} 
    < a_{k l_1} a_{k' l_1'} a_{k' l_2'} a_{k l_2}
       e^{ - i ( \delta_{k l_1}  - \delta_{k' l_1'} 
               + \delta_{k' l_2'} - \delta_{k l_2} ) } >  \times \nonumber \\
  & & {\ } {\ } 
        (\bar{E}_{l_1} - \bar{E}_{l_1'}) (\bar{E}_{l_2'} - \bar{E}_{l_2})
         \times \nonumber  \\
  & & {\ } {\ } \int d^3x \; \Phi_{l_1}^{\dag}(x)\nabla \Phi_{l_1'}(x)\cdot
  \int d^3x \; \Phi_{l_2'}^{\dag}(x)\nabla \Phi_{l_2}(x)  \nonumber \\
  & & =  \sum_{l, l'}  < a_{k l}^2 > <a_{k' l'}^2 >
   2 \epsilon_0^2 |t_c|^2 \left[ \delta ( l' - l -1)
  +  \delta ( l' - l +1) \right] \; .
\eea
Here $t_c = k_F/2$.
In  dirty limit, all core states of $\bar{\cal H}$ are combined to 
make the core states of ${\cal H}$,
that is, there is a uniform distribution for $a_{kl}$
in terms of indices of both $l$ and $k$.
The normalization condition for the wavefunction $\Psi_k$ requires
then 
$ < a_{kl}^2 > = 1/N_c $, 
with the total number of core levels 
$  
  N_c \approx 2 ({\Delta_\infty}/{\epsilon_0} ) ({k_F L}/{2\pi}) 
$,
same as in  clean superconductors.
This is implied by the Anderson theorem that there is no change 
in the number of extended states as the superconductor
evolves from clean to dirty limit.
Hence we can treat the core states and extended states separately,
and ignore the mixing between those two groups of states.
The average value of the transition element is now
\[
  < \left| \int d^3x \; \Psi_k^{\dag}(x)\left( \nabla_v {\cal H} \right) 
       \Psi_{k'}(x) \right|^2 >
  = \frac{4 \epsilon_0^2 |t_c|^2}{N_c} \; ,
\]
and the vortex friction coefficient is then
\be
   \eta =  \frac{\pi}{2} {\hbar }
   \left( \frac{k_F L }{2\pi \epsilon_0 }\right)^2
 \frac{4 \epsilon_0^2 |t_c|^2}{N_c}
 = \frac{3\pi^2 }{8}  \hbar  n_e \frac{\Delta_{\infty}}{E_F}  L  \; .
\ee
This result is similar to the value obtained in Ref.\cite{bs}. Hence 
its microscopic justification  has been provided
for the first time.
The magnitude of this vortex friction contribution 
is smaller than the transverse force by a 
factor of $\Delta_{\infty}/E_F$. 

In the above derivation, 
we have ignored the localization effect which depresses the density of 
state, or the superfluid density. 
We justify our assumption here.
There are three energy scales involved in the derivation of vortex dynamics, 
the Fermi energy $E_F$, the energy gap $\Delta_\infty$, and
the core level spacing $\Delta_\infty^2/ E_F$.
The effect of impurities on vortex dynamics is believed appreciable 
 at  $\tau_{tr} \Delta_\infty^2/\hbar E_F \leq 1 $\cite{russian}, 
and the equal of above Eq.(14) to (16)  suggests 
that the impurity starts to be effective at
$\tau_{tr} \Delta_{\infty}/\hbar \, (\Delta_{\infty}/E_F )^2 \sim 1 $.
They indicate that the impurity effect on vortex friction occurs at a rather
weak level, determined by the smallest energy scale in the problem.  
The dirty limit is given by 
 $ \Delta_\infty/E_F < \tau_{tr}  \Delta_{\infty}/\hbar < 1$.
The localization effect is only pronounce in the extremely dirty limit,
the localization regime,
when $\tau_{tr} E_F/\hbar \leq 1$.\cite{ma}
Because $\Delta_\infty/E_F << 1$, away from the localization regime 
the depression of density is indeed negligible.
The un-depressed electronic density applies.

In conclusion, we have 
found that 
the impurity average smoothes the core state energy spectrum, 
a process
necessary for vortex friction.
We have obtained 
that the core states contribution to vortex friction 
is proportional to the normal state resistivity in weak impurity limit
and  saturates in dirty limit.

\noindent Acknowledgments

This work was financially supported by the Swedish Natural Science Research
Council (NFR).

\end{document}